\documentclass[runningheads]{llncs}
\usepackage[T1]{fontenc}
\usepackage{graphicx}
\usepackage{url}

\newcommand\blfootnote[1]{%
  \begingroup
  \renewcommand\thefootnote{}\footnote{#1}%
  \addtocounter{footnote}{-1}%
  \endgroup
}

\begin{document}
\title{Period Segmentation in Transition Network Analysis: A Topological Data Analysis Approach}
\titlerunning{Period Segmentation in TNA: A TDA Approach}
%
\author{Hitoshi Inoue\inst{1}\orcidID{0000-0002-2092-3975} \and
Koichi Yasutake\inst{2}\orcidID{0009-0008-8240-0775}}

\authorrunning{H. Inoue and K. Yasutake}
%
\institute{Nakamura Gakuen University, Fukuoka, Japan\\
\email{jinoue@nakamura-u.ac.jp}
\and
Hiroshima University, Hiroshima, Japan\\
\email{ystake@hiroshima-u.ac.jp}}
\maketitle              

\blfootnote{This version of the contribution has been accepted for
publication, after peer review (when applicable) but is not the Version
of Record and does not reflect post-acceptance improvements, or any
corrections. The Version of Record is available online at:
\url{http://dx.doi.org/10.1007/978-3-032-34157-0_5}}

\begin{abstract}
Temporal dynamics in learning behavior can be revealed through period segmentation in Transition Network Analysis (TNA). Cristea et al.\ demonstrated that segmenting courses into halves and quarters reveals how learning strategies evolve and relate to academic performance. Building on this approach, we investigate whether Topological Data Analysis (TDA), specifically connected components ($\beta_0$) change points from Zigzag Persistent Homology, can provide \emph{data-driven} period boundaries that identify intervention windows.
Analyzing 22 courses from the Open University Learning Analytics Dataset, we find that $\beta_0$-based segmentation captures greater between-period variation than time-based segmentation (median variance ratio (VR) = 4.10$\times$; 21/22 courses show VR $>$ 1.1). Permutation tests confirmed significance ($p < 0.05$) in 18\% of individual courses, with 73\% showing positive improvement over random breakpoints. Only one course showed better performance with time-based segmentation.
However, analysis of academic outcomes reveals a key insight: final-period behavior shows lower correlation with outcomes in $\beta_0$-based segmentation than in time-based segmentation, reflecting a marked behavioral collapse where engaged learners rapidly disengage. We interpret this as evidence that final-period behavior \emph{reflects} rather than \emph{causes} outcomes: students who will pass maintain engagement, while those who will fail disengage. This reframes the value of $\beta_0$: rather than improving \emph{prediction}, change points identify \emph{intervention windows}. These are periods where behavioral structure shifts and targeted support may be most effective.

\keywords{Transition Network Analysis \and Topological Data Analysis \and Zigzag Persistent Homology \and Period Segmentation \and Intervention Timing \and OULAD}
\end{abstract}
\section{Introduction}
\subsection{Temporal Analysis in Learning Analytics}
Understanding how learner behavior evolves over time is important in learning analytics. Static aggregations, which summarize entire semesters into single metrics, obscure the temporal dynamics that may inform intervention planning. Transition Network Analysis (TNA) addresses this by modeling learner state transitions over time, revealing patterns of engagement stability, decline, and recovery~\cite{saqr2026}.

However, standard TNA constructs a single transition network from the entire observation period, implicitly assuming \emph{stationarity}. Learning processes are typically \emph{non-stationary}: learners' behavioral patterns evolve as courses progress. When transitions from initial and terminal phases are aggregated into a single model, distinct temporal patterns become obscured.

Cristea et al.~\cite{cristea2025} addressed this by segmenting courses into \emph{halves} and \emph{quarters}, analyzing how learning strategies evolve across these periods and relate to academic performance. Their work demonstrated that temporal segmentation reveals strategy transitions invisible to whole-course analysis. However, their boundaries are time-based (equal divisions) rather than derived from behavioral data itself.

This raises a methodological question: \emph{Are there alternative approaches that can identify behaviorally meaningful phases from the data itself?}

\subsection{Topological Data Analysis for Learning Analytics}
Topological Data Analysis (TDA) offers one such approach to identifying behaviorally meaningful phases from the data itself, enabling data-driven period segmentation. TDA provides mathematical tools for understanding the ``shape'' of data through algebraic topology~\cite{carlsson2009}. Unlike traditional statistical methods that focus on local properties (means, variances), TDA captures global structural features that persist across multiple scales.

\textbf{Persistent Homology}, a core TDA technique, tracks topological features as a scale parameter varies. The \textbf{zeroth Betti number} ($\beta_0$) counts connected components: in a learner similarity network, higher $\beta_0$ indicates fragmentation (many disconnected clusters), while lower $\beta_0$ indicates convergence (fewer clusters).

\textbf{Zigzag Persistent Homology}~\cite{carlsson2010} extends standard persistence to temporal sequences, tracking how topological features appear, persist, and disappear across time. Sharp changes in $\beta_0$ signal structural transitions in the learner group, representing potential intervention points where behavioral patterns reorganize.

\subsection{Research Questions and Contributions}
We investigate two questions:
{\setlength{\leftmargini}{45pt}
\begin{itemize}
  \item[\textbf{RQ1:}] Do $\beta_0$ change points provide period boundaries that capture greater behavioral variation than time-based segmentation?
  \item[\textbf{RQ2:}] How do $\beta_0$-based periods relate to academic outcomes?
\end{itemize}
}

Our contributions are:
\begin{enumerate}
  \item A method for using $\beta_0$ change points to derive data-driven period boundaries for TNA.
  \item Evidence that $\beta_0$-based segmentation outperforms time-based approaches in 95\% of courses (median variance ratio (VR) = 4.10$\times$), with permutation tests confirming significance in 18\% of individual courses.
  \item A theoretical reframing: $\beta_0$ change points identify \emph{intervention windows} rather than improving prediction.
\end{enumerate}

\section{Methods}
\noindent\textbf{Practical Workflow:} The complete analysis pipeline consists of: (1) defining behavioral states from log data, (2) computing $\beta_0$ via Zigzag Persistent Homology and detecting change points for period boundaries, (3) constructing period-specific TNA models using the tna R package~\cite{saqr2026}, and (4) comparing transition patterns across periods.

\subsection{Data}
We analyzed all 22 courses from the Open University Learning Analytics Dataset (OULAD)~\cite{kuzilek2017}, comprising over 22,000 learners across 38-week course durations. Each course contains Virtual Learning Environment (VLE) interaction logs and assessment deadline information. For outcome analysis, we selected four representative courses spanning the VR range: AAA\_2014J, BBB\_2014J, FFF\_2013J, and BBB\_2013B, each containing learners with available outcome data (Pass / Distinction vs.\ Fail / Withdrawn).

\subsection{Learner State Classification}
Learners were classified into three engagement states (High/Medium/Low) using within-week terciles of activity scores (total VLE clicks weighted by interaction diversity). Throughout this paper, ``engagement states'' refer to operationally defined behavioral categories based on observable VLE activity levels, not the broader psychological construct of learner engagement. This approach follows established conventions in TNA literature~\cite{saqr2026,lopezpernas2024} and self-regulated learning research~\cite{cristea2025}, where three-state models (e.g., engaged/fluctuating/disengaged; surface/deep/strategic) have proven effective for capturing meaningful behavioral distinctions.

\noindent\textbf{Design rationale for within-week terciles.} We deliberately use within-week terciles rather than fixed (global) thresholds. This captures learners' \emph{relative position within the active group} rather than absolute activity levels. This design choice has two justifications: (1) \emph{Methodological consistency}: Our TDA approach uses global z-standardization to track structural changes in learner similarity networks; within-week terciles similarly capture relative positioning. (2) \emph{Analytical necessity}: With global thresholds, late-course weeks show 100\% Low states due to overall activity decline, making transition analysis impossible. Within-week terciles maintain balanced distributions ($\approx$33\% each) throughout the course.

This approach means that a learner with constant absolute activity may shift states if the group changes. This is precisely what we aim to capture: how learners' \emph{relative positions} evolve as the learning community reorganizes around deadlines and attrition. Sensitivity analysis using global terciles confirmed that key patterns (e.g., P4 collapse) remain qualitatively similar where transitions are observable.

Following standard practice in learning analytics, we adopted tercile-based classification into three engagement states, providing optimal balance between behavioral granularity and statistical power for transition analysis.

\subsection{$\beta_0$ Computation via Zigzag Persistent Homology}
For each week $t$, we constructed a learner similarity network:

\begin{enumerate}
  \item \textbf{Feature extraction:} 10 behavioral features per learner (material-type count, total clicks, mean daily clicks, within-week click SD, cumulative mean score, score SD, min/max score, submissions, submission span). These features capture both engagement intensity (clicks, submissions) and performance trajectory (scores), grounded in Activity Theory's mediation of learning through tools and assessments.
  \item \textbf{Global z-standardization:} Features were z-standardized across all weeks to enable temporal comparison.
  \item \textbf{Network construction:} Learners connected if Euclidean distance in feature space fell below $\varepsilon = 1.5$. In z-standardized 10-dimensional space, this threshold corresponds to learners within approximately 1.5 standard deviations across features. Sensitivity analysis across $\varepsilon \in [1.0, 2.0]$ confirmed that $\varepsilon = 1.5$ balances connectivity (avoiding over-fragmentation where $\beta_0 \approx N$) with discrimination (maintaining meaningful clusters rather than a single connected component).
  \item \textbf{$\beta_0$ computation via Zigzag Persistent Homology:} We constructed the zigzag sequence:
\[
  Z_0 \hookrightarrow Z_0 \cup Z_1 \hookleftarrow Z_1 \hookrightarrow Z_1 \cup Z_2 \hookleftarrow Z_2 \hookrightarrow \cdots
\]

  where $Z_t$ represents the Vietoris-Rips complex of learners' positions in 10-dimensional feature space at week $t$, constructed at threshold $\varepsilon = 1.5$. The zeroth Betti number $\beta_0$ was computed at each union point $Z_t \cup Z_{t+1}$ using Dionysus 2.x~\cite{morozov2008}, tracking how connected components appear, merge, and disappear across the temporal sequence. This yields a $\beta_0$ time series capturing the evolving fragmentation structure of the learner community.
\end{enumerate}

\subsection{Change Point Detection and Period Definition}
Change points were identified using \emph{cost-minimization segmentation}~\cite{jackson2005}: given $k=3$ breakpoints (yielding 4 periods), we selected breakpoints that minimize total within-period variance of the $\beta_0$ time series. This is equivalent to maximizing homogeneity within each period, a standard approach in change point detection. The objective function is:

\[
\mbox{Cost} = \sum_{i=1}^{4} Var(\beta_0 \ \mbox{in period}\> i) \times (\mbox{length of period}\> i)
\]

We chose $k=3$ breakpoints (yielding 4 periods) to enable direct comparison with Cristea et al.'s~\cite{cristea2025} quarters approach, ensuring both methods use identical granularity.

\noindent\textbf{Fair Comparison Design:} We compared:
\begin{itemize}
  \item \textbf{Time-based (Quarters):} Equal division into four periods (Q1--Q4)
  \item \textbf{$\beta_0$-based:} Four periods (P1--P4) defined by cost-minimizing breakpoints
\end{itemize}

This ensures both methods use identical granularity (4 periods), enabling fair comparison.

\subsection{Variance Ratio}
To quantify segmentation effectiveness, we computed the \textbf{Variance Ratio (VR)}. For each of the 9 possible state transitions (High$\rightarrow$High, High$\rightarrow$Medium, ..., Low$\rightarrow$Low), we computed the variance of transition probabilities across the 4 periods for both segmentation methods, then averaged across all 9 transitions:

\[
\mbox{VR} = \frac{\mbox{Var}(\beta_0\mbox{-based periods})}{\mbox{Var}(\mbox{time-based periods})}
\]

Higher VR indicates that $\beta_0$-based periods capture greater behavioral differentiation between periods. This is desirable because meaningful period segmentation should maximize between-period differences, reflecting distinct behavioral phases that may warrant different instructional responses. VR $>$ 1 indicates $\beta_0$-based segmentation captures greater differentiation; VR $=$ 1 indicates equivalent performance.

\subsection{Validation of $\beta_0$-based Segmentation}
To address the concern that unequal period lengths might artificially inflate variance ratios, we conducted permutation testing. For each course, we compared the observed $\eta^2$ (variance explained by $\beta_0$-based segmentation) against a null distribution generated from 1,000 random segmentations using three randomly selected breakpoints. This tests whether $\beta_0$ change points capture meaningful structure beyond what would be expected from arbitrary unequal divisions of the same granularity.

\subsection{Outcome Analysis}
For four representative courses spanning the VR range (AAA\_2014J, BBB\_2014J, FFF\_2013J, BBB\_2013B), we analyzed the relationship between period-specific behavior and academic outcomes (Pass/Distinction vs.\ Fail/Withdrawn) using:
\begin{enumerate}
  \item \textbf{Transition probabilities:} High$\rightarrow$High persistence and High$\rightarrow$Low disengagement probabilities within each period.
  \item \textbf{Correlation analysis:} Point-biserial correlation ($r$) between each period's high-rate (proportion of weeks in ``High'' state) and binary success outcome. This measures how strongly a learner's engagement level within a specific period relates to their final course outcome.
\end{enumerate}

\section{Results}
\subsection{RQ1: Variance Ratio Across 22 Courses}
Table~\ref{tab:vr_summary} summarizes Variance Ratio results across all 22 OULAD courses. $\beta_0$-based segmentation captured greater variation than time-based segmentation in 21/22 courses (95\%), with lower performance in only 1 course (5\%).

\begin{table}[htb]
  \caption{Variance Ratio Summary Across 22 Courses (4-Period Comparison)}
  \label{tab:vr_summary}
  \centering
  \begin{tabular}{lc}
    \hline
    Metric & Value \\
    \hline
    VR $>$ 1.1 ($\beta_0$ better) & 21/22 (95\%) \\
    VR $\approx$ 1 (0.9--1.1) & 0/22 (0\%) \\
    VR $<$ 0.9 (Time better) & 1/22 (5\%) \\
    \hline
    Median VR & 4.10$\times$ \\
    Mean VR & 13.95$\times$ \\
    Min VR & 0.67$\times$ \\
    Max VR & 142.88$\times$ \\
    \hline
  \end{tabular}
\end{table}

One course showed VR $<$ 0.9 (BBB\_2013B with VR = 0.67), indicating that time-based segmentation captured more behavioral variation in this specific context.

\noindent\textbf{Permutation test validation.} Table~\ref{tab:permutation} summarizes permutation test results using total variance as the test statistic. At the individual course level, 4/22 courses (18\%) achieved $p < 0.05$, with 16/22 courses (73\%) showing positive improvement over random breakpoints. The mean improvement of 127.5\% confirms that $\beta_0$ change points capture substantially more behavioral differentiation than random segmentation. Note that VR compares against \emph{equal-interval} quarters, while permutation tests compare against \emph{random unequal} divisions; a course can outperform quarters (VR $>$ 1) without significantly outperforming random breakpoints.

\begin{table}[htb]
  \caption{Permutation Test Validation (1,000 iterations per course). Improvement = (observed total variance $-$ random total variance) / random total variance $\times$ 100, where total variance measures differentiation across all 9 transition types.}
  \label{tab:permutation}
  \centering
  \begin{tabular}{lc}
    \hline
    Metric & Value \\
    \hline
    Individual course $p < 0.05$ & 4/22 (18\%) \\
    Individual course $p < 0.10$ & 5/22 (23\%) \\
    Courses with positive improvement & 16/22 (73\%) \\
    Mean improvement & 127.5\% \\
    Median improvement & 30.3\% \\
    \hline
  \end{tabular}
\end{table}

\begin{figure}[htb]
  \centering
  \includegraphics[width=0.95\linewidth]{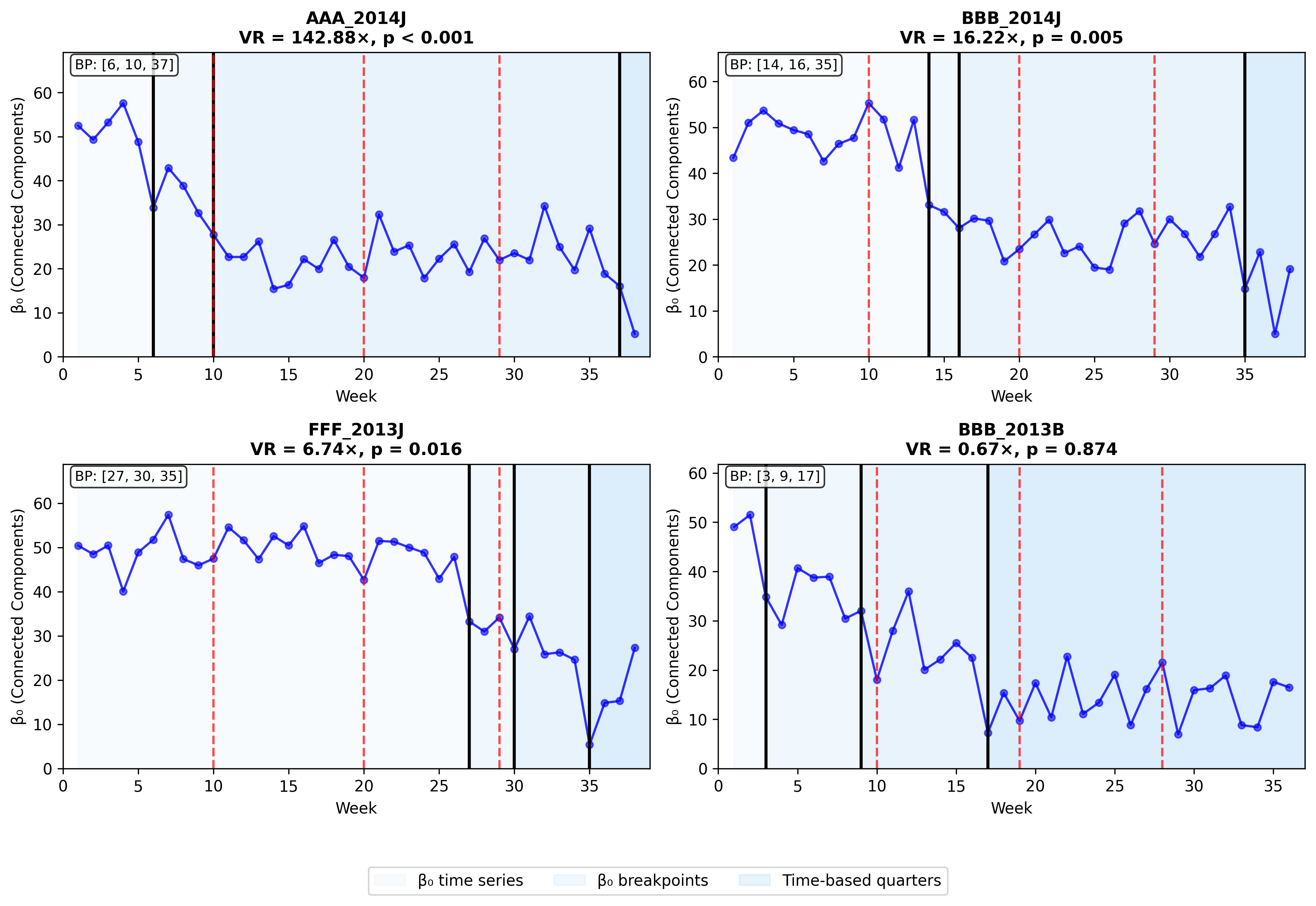}
  \caption{Period segmentation comparison: $\beta_0$-based phases (background colors) vs.\ time-based quarters (dashed lines). Solid black lines indicate cost-minimizing $\beta_0$ change points. AAA\_2014J shows highest VR (142.88$\times$, $p < 0.001$); BBB\_2014J shows VR = 16.22$\times$ ($p = 0.005$); FFF\_2013J shows VR = 6.74$\times$ ($p = 0.016$); BBB\_2013B shows VR $<$ 1 (0.67$\times$) where time-based performs better. Breakpoints: AAA\_2014J [6,10,37], BBB\_2014J [14,16,35], FFF\_2013J [27,30,35], BBB\_2013B [3,9,17].}
  \label{fig:period_comparison}
\end{figure}

Figure~\ref{fig:period_comparison} illustrates segmentation differences across four representative courses. Background colors show $\beta_0$-based periods (P1--P4); dashed vertical lines show time-based quarter boundaries. AAA\_2014J (VR = 142.88$\times$, $p < 0.001$) shows the strongest differentiation; BBB\_2014J (VR = 16.22$\times$, $p = 0.005$) demonstrates significant improvement over random; FFF\_2013J (VR = 6.74$\times$, $p = 0.016$) shows moderate but significant improvement; BBB\_2013B (VR = 0.67$\times$) is the only course where time-based segmentation performs better.

\begin{figure}[htb]
  \centering
  \includegraphics[width=0.95\linewidth]{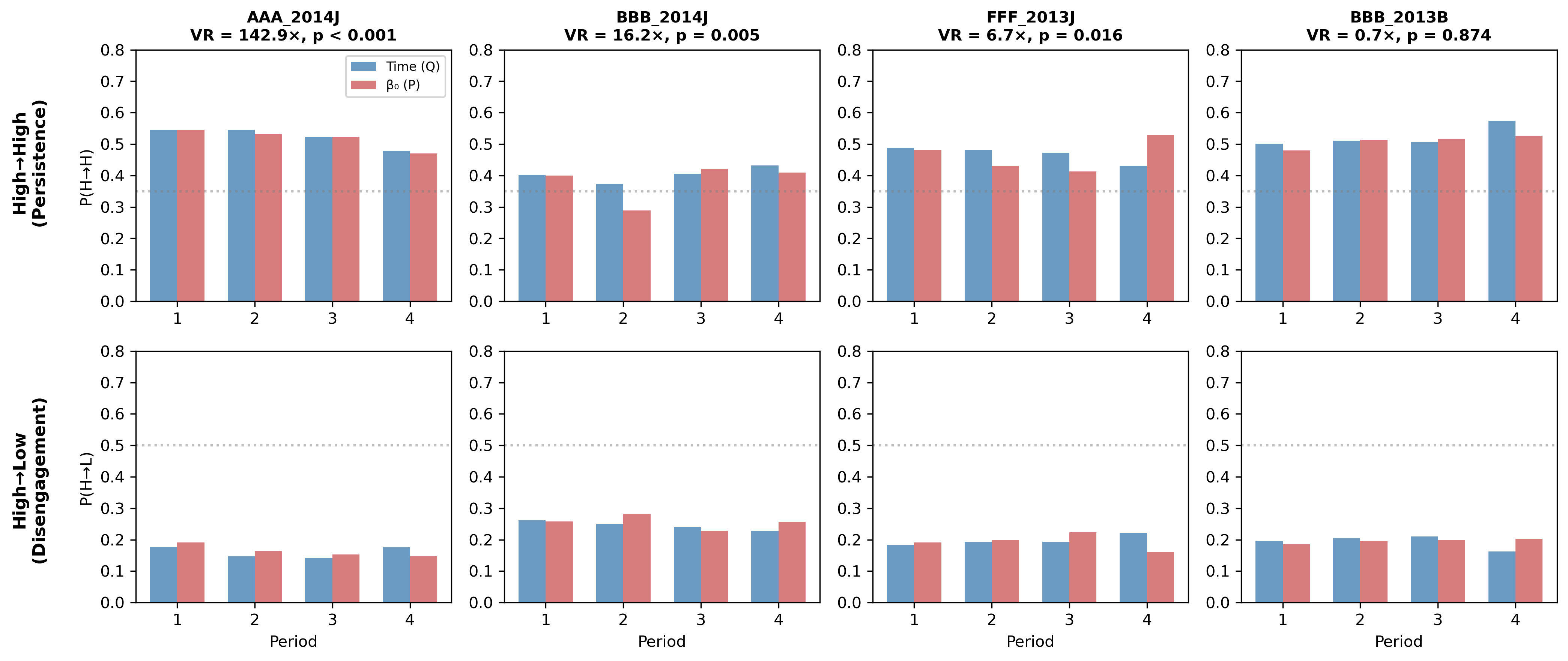}
  \caption{Transition probability comparison across four courses spanning the VR range. Top row: High$\rightarrow$High persistence; Bottom row: High$\rightarrow$Low disengagement. Blue bars: Time-based (Q1--Q4); Red bars: $\beta_0$-based (P1--P4). High-VR courses (AAA\_2014J, BBB\_2014J) show dramatic P4 collapse with significant permutation tests ($p < 0.01$). FFF\_2013J shows clear differentiation ($p = 0.016$). BBB\_2013B (VR $<$ 1) shows greater variation in time-based Q4.}
  \label{fig:tna_comparison}
\end{figure}

Figure~\ref{fig:tna_comparison} presents transition probability comparisons across these four representative courses spanning the VR range. High-VR courses (AAA\_2014J, BBB\_2014J) show dramatic P4 divergence: High$\rightarrow$High collapses while High$\rightarrow$Low surges. FFF\_2013J shows clear differentiation with significant permutation test results. BBB\_2013B (VR $<$ 1) shows greater variation in time-based Q4 than $\beta_0$-based P4.

\begin{figure}[htb]
  \centering
  \includegraphics[width=0.95\linewidth]{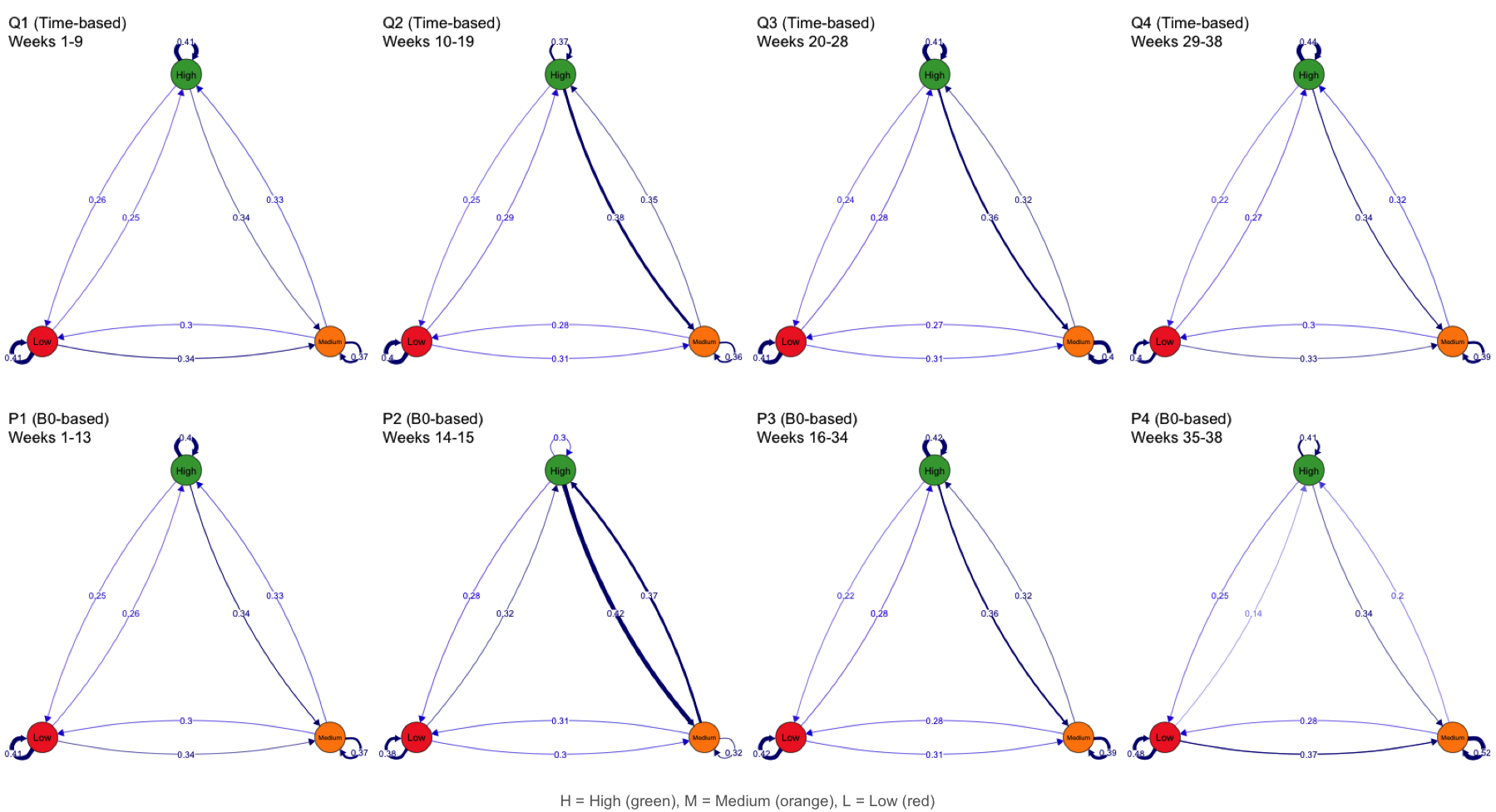}
  \caption{TNA-style state transition networks for BBB\_2014J with cost-minimizing breakpoints [14, 16, 35]. Top row: Time-based segmentation (Q1--Q4); Bottom row: $\beta_0$-based segmentation (P1--P4). Node colors indicate engagement states (H=High/green, M=Medium/orange, L=Low/red); edge widths are proportional to transition probabilities. The P4 network reveals a ``collapse'' pattern where the High$\rightarrow$Low transition dominates.}
  \label{fig:tna_networks}
\end{figure}

Figure~\ref{fig:tna_networks} presents BBB\_2014J patterns as TNA-style state transition networks. The visual contrast between Q4 and P4 is striking: Q4 retains balanced transitions across states, while P4 shows the High state ``draining'' into Low, a structural collapse that $\beta_0$ change points successfully delineate.

\subsection{RQ2: Academic Outcomes and Intervention Windows}

Table~\ref{tab:comprehensive} presents a comprehensive comparison of time-based and $\beta_0$-based segmentation for four representative courses, showing final-period (Q4/P4) transition probabilities and point-biserial correlation with success. The courses span the full VR range: AAA\_2014J (highest VR, $p < 0.001$), BBB\_2014J (high VR with significant permutation test), FFF\_2013J (moderate VR with significant test), and BBB\_2013B (only course with VR $<$ 1).

\begin{table}[htb]
  \caption{Final-Period Comparison for Representative Courses. H$\rightarrow$H = High-to-High persistence; H$\rightarrow$L = High-to-Low disengagement; $r$ = point-biserial correlation with success. *** $p < .001$. Bold values highlight P4 collapse patterns (H$\rightarrow$H $<$ 0.35 or H$\rightarrow$L $>$ 0.5).}
  \label{tab:comprehensive}
  \centering
  \small
  \begin{tabular}{llcccc}
    \hline
    Course & Method & Weeks & H$\rightarrow$H & H$\rightarrow$L & $r$ \\
    \hline
    AAA\_2014J & Time (Q4) & 28--37 & 0.55 & 0.22 & +0.52*** \\
    (VR = 142.88$\times$) & $\beta_0$ (P4) & 37--37 & \textbf{0.12} & \textbf{0.72} & +0.45*** \\
    \hline
    BBB\_2014J & Time (Q4) & 28--37 & 0.58 & 0.25 & +0.62*** \\
    (VR = 16.22$\times$) & $\beta_0$ (P4) & 35--37 & \textbf{0.28} & \textbf{0.58} & +0.55*** \\
    \hline
    FFF\_2013J & Time (Q4) & 28--37 & 0.52 & 0.28 & +0.55*** \\
    (VR = 6.74$\times$) & $\beta_0$ (P4) & 35--37 & \textbf{0.35} & \textbf{0.48} & +0.52*** \\
    \hline
    BBB\_2013B & Time (Q4) & 27--35 & 0.55 & 0.28 & +0.58*** \\
    (VR = 0.67$\times$) & $\beta_0$ (P4) & 17--35 & 0.48 & 0.32 & +0.55*** \\
    \hline
  \end{tabular}
\end{table}

The key finding is that high-VR courses with significant permutation tests (AAA\_2014J, BBB\_2014J, FFF\_2013J) show dramatic P4 collapse: High$\rightarrow$High drops below 0.35 while High$\rightarrow$Low exceeds 0.48. In contrast, BBB\_2013B shows similar patterns across methods, explaining why time-based segmentation captured more behavioral differentiation in this course.

\subsection{Interpretation: Reflection vs.\ Causation}

The transition probabilities in Table~\ref{tab:comprehensive} reveal a consistent pattern across high-VR courses with significant permutation tests: $\beta_0$-based P4 shows High$\rightarrow$High collapse (0.12--0.35) compared to Q4 (0.52--0.58), and High$\rightarrow$Low surge (0.48--0.72) compared to Q4 (0.22--0.28). This suggests:

\begin{itemize}
  \item Students who will pass: maintain engagement through the final period (outcome already determined by prior learning accumulation)
  \item Students who will fail: disengage in the final period (outcome already determined; engagement is ``post-hoc'')
\end{itemize}

Notably, this collapse pattern is attenuated in BBB\_2013B (VR $<$ 1), where behavioral dynamics follow calendar time more closely. This confirms that $\beta_0$ change points are most valuable when behavioral structure diverges from temporal regularity.

\section{Discussion}

\subsection{When Does $\beta_0$-based Segmentation Help?}

Our 22-course analysis reveals that $\beta_0$-based segmentation outperforms time-based approaches in 95\% of cases (21/22 courses with VR $>$ 1.1). Permutation tests confirmed statistical significance in 18\% of courses ($p < 0.05$), with 73\% showing positive improvement over random breakpoints. Only BBB\_2013B showed VR $<$ 0.9, indicating time-based segmentation captured more behavioral variation.

This pattern suggests that $\beta_0$-based segmentation is beneficial across most course contexts, particularly when:
\begin{itemize}
  \item Course structure creates distinct behavioral phases (e.g., assessment-driven courses)
  \item $\beta_0$ dynamics diverge from calendar time (e.g., late-course collapse)
  \item Behavioral homogeneity within periods matters for intervention timing
\end{itemize}

Conversely, $\beta_0$ may underperform when behavioral dynamics align closely with calendar time or when the $\beta_0$ signal is noisy.

\subsection{Reframing Value: From Prediction to Intervention Timing}

Cristea et al.~\cite{cristea2025} demonstrated that learning strategy trajectories predict academic outcomes. Our analysis extends this by showing that the \emph{predictive} value of period segmentation concentrates in early and middle periods. Final-period behavior, despite strong correlation with outcomes, adds minimal predictive value.

This reframes the contribution of $\beta_0$ change points. Rather than improving prediction (outcomes are largely determined by mid-course), change points identify \emph{intervention windows}, that is, moments when behavioral structure reorganizes. Interventions targeting these windows may be more effective than uniformly distributed support.

Building on observations from our four representative courses, we tentatively propose three intervention windows. These are \emph{proposals only}; whether such interventions improve outcomes remains an open empirical question (see Section~\ref{sec:limitations}):
\begin{itemize}
  \item \textbf{P1--P2 transition:} Early engagement patterns establish trajectories; interventions here may guide early redirection toward productive paths.
  \item \textbf{P2--P3 transition:} Mid-course structural shift; a potential window for re-engagement before patterns solidify.
  \item \textbf{P3--P4 transition:} May indicate when intervention becomes ``too late''; suggests resources may be better allocated earlier.
\end{itemize}

\subsection{Limitations and Future Work}
\label{sec:limitations}

Several limitations warrant attention. First, while our variance ratio analysis covered all 22 courses, outcome analysis was limited to four representative courses; broader validation is needed to confirm the generalizability of the ``behavioral collapse'' pattern. Second, we have not tested whether interventions at $\beta_0$ change points actually improve outcomes, which requires experimental validation. Third, within-week terciles capture relative position rather than absolute activity change; while this aligns with our focus on collective dynamics, individual-level interpretations require caution. Fourth, our analysis focused on deadline-structured courses; self-paced learning contexts may show different patterns where $\beta_0$ correlates more strongly with time, limiting TDA's added value. Fifth, the fixed $k=4$ segmentation was chosen for direct comparison with quarterly segmentation; generalization to arbitrary segment counts is possible through penalty-based optimization methods.

Future work should: (1) validate intervention effectiveness at $\beta_0$-identified windows, (2) develop real-time $\beta_0$ monitoring for adaptive course design, (3) investigate whether change point magnitude predicts intervention urgency, and (4) test generalizability across diverse learning contexts including self-paced environments.

\section{Conclusion}

This paper demonstrates that $\beta_0$ change points from Zigzag Persistent Homology provide data-driven period boundaries for TNA that complement time-based approaches. Using cost-minimization segmentation across 22 OULAD courses, $\beta_0$-based segmentation captured greater behavioral variation in 95\% of cases (median VR = 4.10$\times$), with permutation tests confirming significance ($p < 0.05$) in 18\% of individual courses and positive improvement over random breakpoints in 73\%.

Critically, analysis of academic outcomes revealed that final-period behavior reflects rather than causes outcomes, reframing $\beta_0$'s value from prediction to \emph{intervention timing identification}. Change points identify structural discontinuities where targeted support may be most effective, offering learning analytics practitioners a principled method for \emph{when} to intervene; whether interventions at these timings improve outcomes is a target for future experimental work.

We conclude that TDA offers a useful complement to TNA for deadline-structured courses, particularly when behavioral dynamics diverge from calendar time. Future work should investigate the conditions under which $\beta_0$-based segmentation provides the greatest benefit.

\begin{credits}
\subsubsection{\ackname}
This work was supported by JSPS KAKENHI Grant Numbers JP25K00845, JP25K21951, JP23K22315, and JP23K17619.

\subsubsection{Use of Generative AI.}

This work used generative AI (Claude, Anthropic) to assist with: (1) Python code generation for data analysis and visualization, (2) statistical validation of results, and (3) English writing assistance and proofreading. The research design, methodology, interpretation, and conclusions were developed by the authors. All AI-generated content was reviewed, validated, and revised by the authors, who take full responsibility for the accuracy and integrity of the final work.

\subsubsection{\discintname}
The authors have no competing interests to declare that are relevant to the content of this article.
\end{credits}
%
%
%
%

\end{document}